\documentclass{emulateapj}
\usepackage{graphics}

\slugcomment{Accepted to The Astrophysical Journal,  Jul 3, 2013}

\newcommand \msun  {M$_{\odot}$}
\newcommand \mic  {$\mu$m}

\begin{document}


\shortauthors{TEMIM \& DWEK}

\shorttitle{DUST HEATING MODELS} 

\title{THE IMPORTANCE OF PHYSICAL MODELS FOR DERIVING DUST MASSES AND GRAIN SIZE DISTRIBUTIONS IN SUPERNOVA EJECTA I: RADIATIVELY HEATED DUST IN THE CRAB NEBULA}

\author{TEA TEMIM\altaffilmark{1,2}, ELI DWEK\altaffilmark{1}}

\altaffiltext{1}{Observational Cosmology Lab, Code 665, NASA Goddard Space Flight Center, Greenbelt, MD 20771, USA}
\altaffiltext{2}{Oak Ridge Associated Universities (ORAU), Oak Ridge, TN  37831, USA; tea.temim@nasa.gov}

\begin{abstract}

Recent far-infrared (IR) observations of supernova remnants (SNRs) have revealed significantly large amounts of newly-condensed dust in their ejecta, comparable to the total mass of available refractory elements. The dust masses derived from these observations assume that all the  grains of a given species radiate at the same temperature, regardless of the dust heating mechanism or grain radius.  In this paper, we derive the dust mass in the ejecta of the Crab Nebula, using a physical model for the heating and radiation from the dust. We adopt a power-law distribution of grain sizes and two different dust compositions (silicates and amorphous carbon), and calculate the heating rate of each dust grain by the radiation from the pulsar wind nebula (PWN). We find that the grains attain a continuous range of temperatures, depending on their size and composition. The total mass derived from the best-fit models to the observed IR spectrum is $0.019-0.13 \rm \: M_{\odot}$, depending on the assumed grain composition. We find that the power-law size distribution of dust grains is characterized by a power-law index of 3.5-4.0 and a maximum grain size larger than 0.1~\mic. The grain sizes and composition are consistent with what is expected  for dust grains formed in a Type IIP SN. 
Our derived dust mass is at least a factor of two less than the mass reported in previous studies of the Crab Nebula that assumed more simplified two-temperature models. 
These models also require a larger mass of refractory elements to be locked up in dust than  was likely available in the ejecta. 
The results of this study show that a physical model resulting in a realistic distribution of dust temperatures can constrain the dust properties and affect the derived dust masses. 
Our study may also have important implications for deriving grain properties and mass estimates in other SNRs and for the ultimate question of whether SNe are major sources of dust in the Galactic interstellar medium (ISM) and in external galaxies.

\end{abstract}


\section{INTRODUCTION} \label{intro}

In recent years, the far-infrared (IR) observations of supernova remnants (SNRs) with the \textit{Herschel Space Observatory} \citep{pil10} have allowed us to finally tackle the longstanding question of whether supernovae (SNe) contribute a significant amount of dust to the interstellar medium (ISM). Theoretical dust condensation models predict that 0.1-0.7 $\rm M_{\odot}$ of dust should form in a SN explosion of a $\sim20\: \rm M_{\odot}$ star. \citep{tod01,che10,noz10,sil12}. A significant fraction of this newly condensed dust may be destroyed following the encounter with the reverse shock in the ejecta \citep[e.g.,][]{dwe05,bia05,koz09}, and SN blast waves in the interstellar mediam \citep[][and references therein]{jon11}. SNe may therefore be required to account for the mass of dust observed in local \citep[e.g.][]{dwe80,mat09,cal11,boy12,zhu13} and high-redshift galaxies \citep[e.g.,][]{dun03,mor03, dwe07,mic11,gall11a,gall11b,dwe11,val12}. Required SN dust yields may be as low as $\sim 0.1$~\msun\ or as high as $\sim 1-2$~\msun, depending on the grain destruction efficiency in the ISM \citep{dwe07}. 

\textit{Spitzer Space Telescope} observations of SNRs have revealed ejecta dust in many remnants with estimated masses in the 0.02--0.1$\rm M_{\odot}$ range \citep{sug06,rho08,rho09,tem10,tem12b}, but the recent far-IR observations that are sensitive to cooler dust are now suggesting even larger masses. Cas A appears to contain $0.075 \rm \: M_{\odot}$ of cool ($\sim$ 35 K) ejecta dust located on interior of the reverse shock \citep{bar10,sib10}, raising the total dust mass to $0.1 \rm \: M_{\odot}$ \citep{rho09}. Based on recent \textit{Herschel} observations, \citet{mat11} reported 0.4-0.7 $\rm M_{\odot}$ of cool dust in SN 1987A, while \citet{gom12} find $0.12-0.25 \rm \: M_{\odot}$ of dust in the Crab Nebula.

While these recent results imply that SNe may indeed be important suppliers of dust, the masses in each each of these cases are derived using 
simple modified blackbody distributions with one or two dust temperature
components to fit the IR and submillimeter spectral energy distribution. A more realistic scenario is to identify the heating source in the ejecta, and use a continuous size distribution of grains with each grain heated to a different temperature that depends on its composition, size and optical properties. For example, \cite{ric13} modeled the $\rm H_2$ emission in the Crab Nebula, and showed that a given grain composition and size distribution leads to continuous distribution of dust temperatures. 

In this work, we present a detailed model of the physical mechanism giving rise to the observed IR emission from the Crab Nebula. 
Dust in the Crab Nebula was discovered as an IR excess above the synchrotron power-law spectrum of its pulsar wind nebula (PWN)\citep{tri77,gla82,mar84,dou01,gre04,tem06}. Absorption features from dust are observed to spatially coincide with the ejecta filaments \citep{wol87,fes90, hes90, bla97,lol10}, and a recent analysis of the \textit{Spitzer} IRS spectra confirmed that the dust is indeed located in the filament cores \citep{tem12b}, and has therefore condensed in the ejecta.

The model compares three different dust compositions, each characterized by a power-law distribution in grain radii whose index is allowed to vary. The primary goal of our study is to determine whether a physical model for the IR emission, characterized by a continuous distribution in dust temperatures will affect the derived grain properties and dust mass in this remnant. Our choice of the Crab Nebula is primarily motivated by the fact that the dust grains are heated by the synchrotron radiation from the PWN, so that their temperature can be accurately derived for each size and composition. Furthermore, since the newly-formed dust has neither been processed by the reverse shock, nor mixed with the ambient medium \citep[for review see][]{hes08,lol10}, the Crab Nebula offers a unique opportunity to study the mass, composition, and size distribution of pristine SN-condensed dust.

In spite of the fact that we concentrate on modeling the IR emission from the Crab Nebula, our results may have broader implications for deriving dust properties and masses from IR observations of SNRs.

\section{DUST HEATING MODEL}\label{model}

\subsection{Heating by the PWN}\label{heat}

\begin{figure}
\epsscale{1.2} \plotone{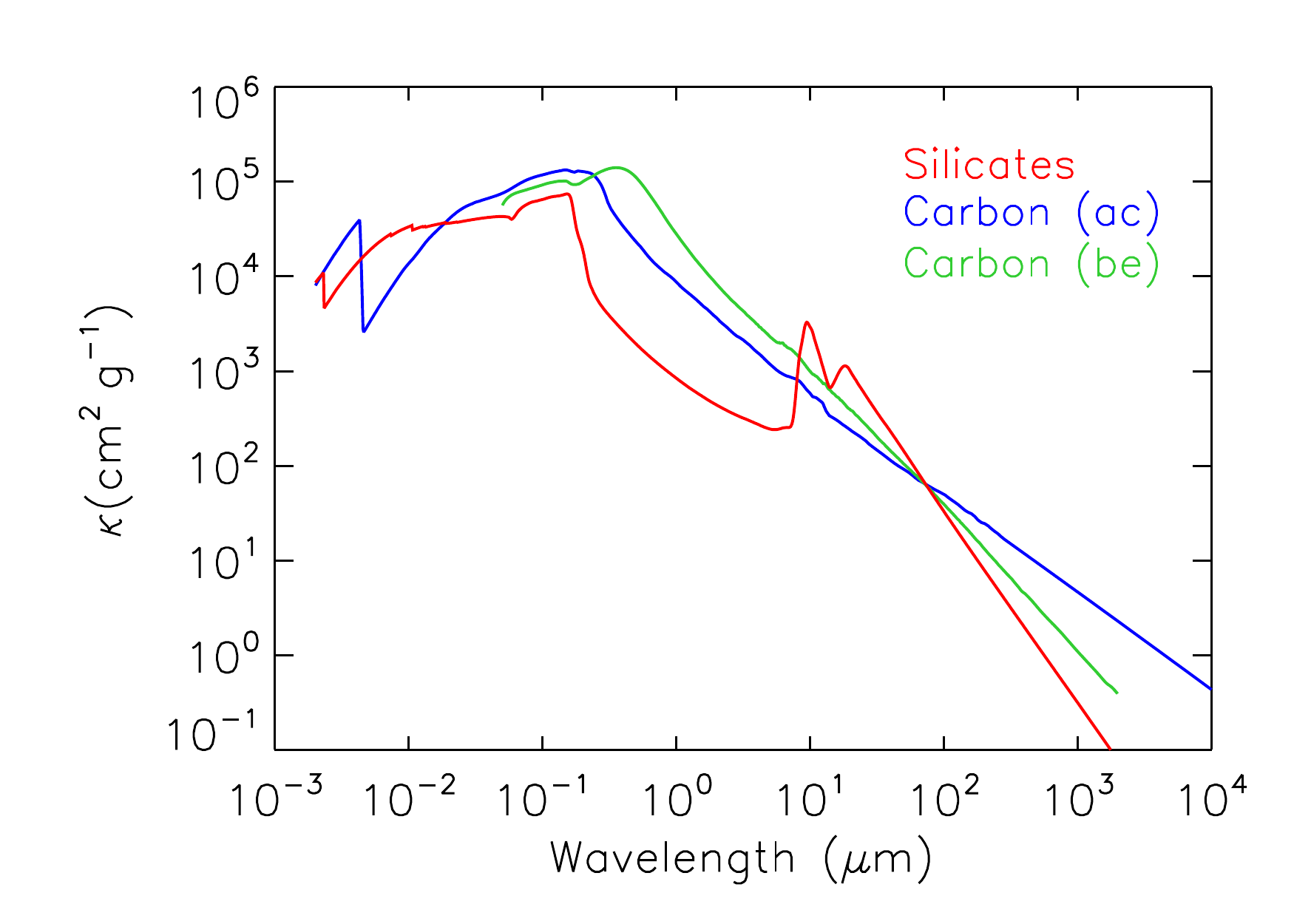}\caption{\label{qabs} Mass absorption coefficients as a function of wavelength where the silicate grains \citep{wei01} are shown in red, amorphous carbon-AC grains \citep{rou91} in blue, and amorphous carbon-BE grains \citep{zub04} in green.}
\end{figure}

Since the dust in the Crab Nebula is concentrated in the Rayleigh-Taylor filaments that form a cage around the PWN \citep{tem12b}, we model the heating of dust by assuming that the heating source is located at the center of the Crab Nebula, at the location of the Crab pulsar. The dominant heating source for the dust in the Crab Nebula is the synchrotron radiation from the PWN, with an insignificant possible contribution from collisional heating by the gas in the filaments \citep{dwe81,tem12b}. In this case, the heating rate (erg~s$^{-1}$) of a single dust grain of radius $a$ is given by 
\begin{equation}\label{heatrate}
H(a)=\frac{\pi a^2 \int L_{\nu}(\nu)Q_{abs}(\nu,a)d\nu}{4\pi r^2},
\end{equation}
where $L_{\nu}$ is the non-thermal specific luminosity of the Crab Nebula's PWN, $a$ is the grain size, $Q_{abs}(\nu,a)$ is the absorption coefficient for a given grain composition, and $r$ is the distance between the radiation source and the dust grain. The non-thermal luminosity $L_{\nu}(\nu)$ of the Crab Nebula's PWN that was used in our model is summarized in Figure 2 of \citet{hes08}. As described in \citet{tem12b}, we consider $L_{\nu}(\nu)$ up to an energy of about 0.6~keV, since we find that the fraction of the energy deposited in the dust at higher energies makes a negligible contribution of less than 1~\%, due to the the combined effects of decreasing of both $L_{\nu}(\nu)$ with energy, and the efficiency of the energy deposited by the photoelectrons in the dust grain \citep{dwe96}.

\begin{figure}
\epsscale{1.2} \plotone{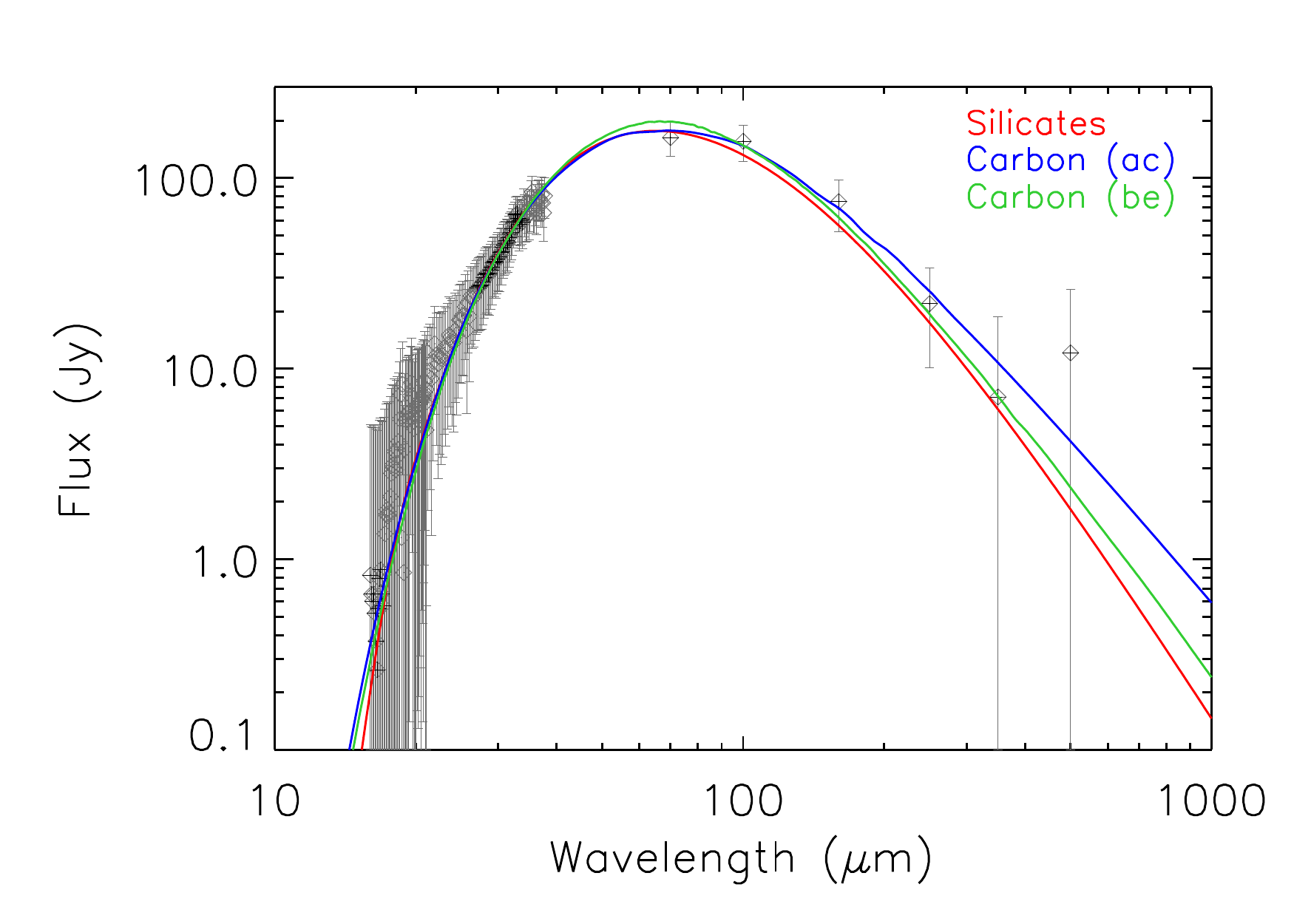}\caption{\label{bestfit} Best-fit dust grain heating models for three different grain compositions (see Figure~\ref{qabs}). The data include the average \textit{Spitzer} IRS dust spectrum \citep{tem12b}, and the dust fluxes from \textit{Herschel} PACS and SPIRE \citep{gom12}. The line-free regions of the IRS spectrum that were used in the fit are shown as black data points. The best-fit parameters are summarized in Table~\ref{dustfitstab}, and corresponding total dust masses in Table~\ref{dustmasstab}.}
\end{figure}

Based on the three-dimensional models of the Crab Nebula, the ejecta filaments are located between 0.55-1.0 pc from the center of the nebula \citep[e.g.][]{cad04}. In our dust model, we allow the distance from the heating source $r$ to vary between 0.5-1.5 pc in intervals of 0.2 pc. We allow the upper limit on the distance to extend beyond the physical location of the filaments in order to account for any attenuation by dust that would affect the heating rate. The best-fit models described in Section~\ref{results} favor the low end of the distance range, suggesting that the internal absorption in the nebula is negligible. In \citet{tem12b}, we find that the optical depth is indeed low, with $\tau \leq 1$.

In equilibrium, the heating rate in Equation~(\ref{heatrate}) of each grain of radius $a$ is equal to the radiative cooling rate given by 
\begin{eqnarray}
L_{gr}(a) & = & 4\pi a^2\int\pi B_{\nu}(\nu,T) Q_{abs}(\nu,a)d\nu, \\ \nonumber
 & = & 4\, m_d(a)\, \int\pi B_{\nu}(\nu,T) \kappa(\nu,a)d\nu,
\end{eqnarray} 
where $B_{\nu}$ is the Planck function, $T$ is the dust grain temperature, $\kappa(\nu,a) = 3 Q_{abs}/4\rho a$ is the mass absorption coefficient, $m_d=4\pi \rho a^3/3$ is the dust mass, and $\rho$ is the mass density of a dust grain. At the wavelengths of the IR emission where $\lambda >> a$ (the Rayleigh regime), $\kappa$ is independent of grain radius.  We used this relationship to compute the temperature for each grain radius and composition, and each distance $r$, and produced a final set of dust emission models by convolving the resulting emission spectrum for each grain size with the size distributions functions described in the previous section. 

\begin{figure}
\epsscale{1.2} \plotone{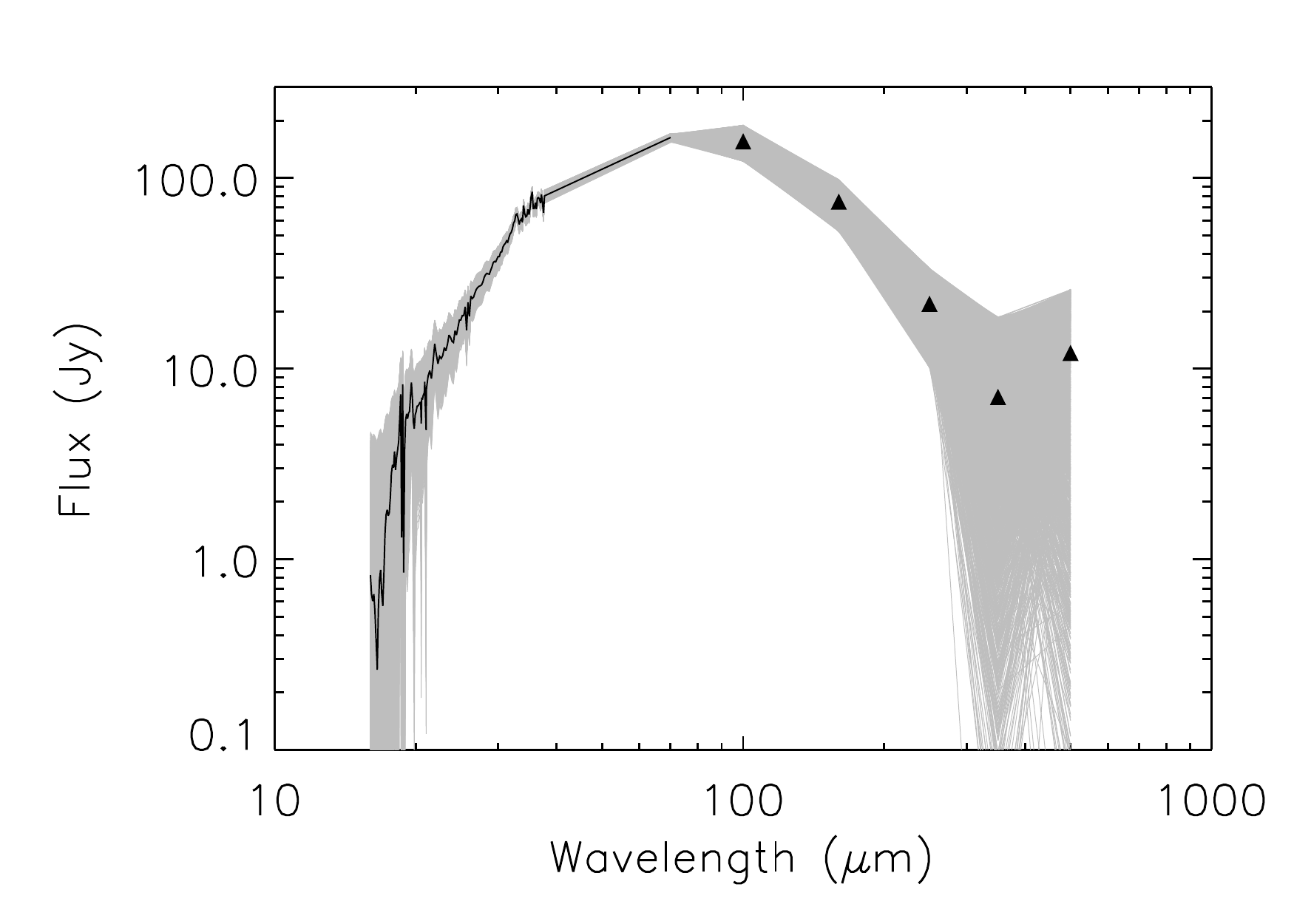}\caption{\label{montecarlospec} The range of spectra simulated by the Monte Carlo  bootstrap method described in Section~\ref{montecarlo} is shown as the gray band, with the actual data overlaid in black.}
\end{figure}

\begin{figure}
\epsscale{1.23} \plotone{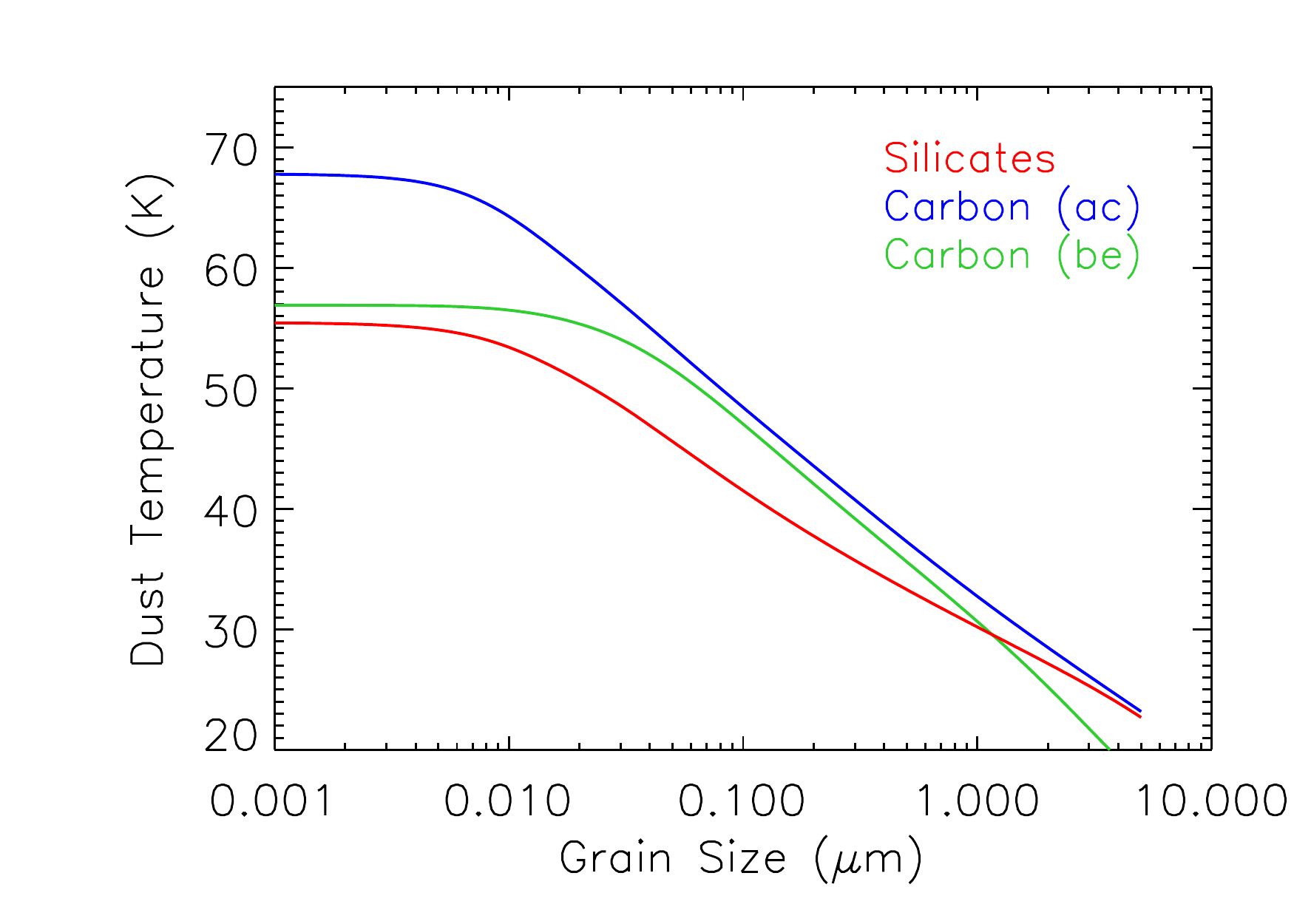}\caption{\label{temp} Grain temperature as a function of grain size for radiatively heated dust in the Crab Nebula, located 0.5 pc from the center of the PWN. The input spectrum of the heating source is the broadband non-thermal spectrum of the Crab Nebula, as summarized in Figure 2 of \citet{hes08}. The colored curves correspond to grains with mass absorption coefficients shown in Figure~\ref{qabs}. As described in Section~\ref{temps}, the large difference in the temperatures of the two carbon compositions is primarily due to the lack of short-wavelength data for the absorptions coefficients of BE carbon.}
\end{figure}

\begin{deluxetable}{lcccc}
\tablecolumns{5} \tablewidth{0pc} \tablecaption{\label{dustfitstab}Best-Fit Model Parameters}
\tablehead{
  \colhead{Composition} &  \colhead{r (pc)}  & \colhead{$\alpha$} &  \colhead{$a_{max}$ (\micron)} & \colhead{$\chi^2$}}
\startdata
Silicates & 0.5 (0.5) & 3.5 (3.5)& 5.0 ($>$ 0.6)& 3.13 \\
Carbon-AC & 0.7 (0.5-0.7)& 4.0 (3.5-4.0)& 0.1 ($>$ 0.1) & 1.86 \\
Carbon-BE & 0.5 (0.5) & 3.5 (3.5) & 0.6 ($>$ 0.3) & 3.14 
\enddata
\tablecomments{Best-fit model parameters where $r$ is the distance of the dust grains from the center of the PWN, $\alpha$ is the power-law index on the grain size distribution, and $a_{max}$ is the maximum grain size cut-off. The values in parentheses represent the range of parameter values obtained from the bootstrap method (see Section \ref{montecarlo}).}
\end{deluxetable}

\begin{deluxetable*}{lccccc}
\tablecolumns{6} \tablewidth{0pc} \tablecaption{\label{dustmasstab}Total Dust Mass ($\rm M_{\odot}$)}
\tablehead{
  \colhead{Composition} &   \multicolumn{2}{c}{Models} & \multicolumn{3}{c}{Nucleosynthesis Yields} \\
      \colhead{} &   \colhead{Physical} &   \colhead{One/Two-Temp} & \colhead{WW95} &  \colhead{N06} & \colhead{WH07}
  }
\startdata
Silicates & 0.13$\pm$0.01 & $0.2\pm0.1$ & 0.08 & 0.32 & 0.12 \\
Carbon-AC & 0.019$^{+0.010}_{-0.003}$ & $0.018\pm0.005$ & 0.05 & 0.10 & 0.11 \\
Carbon-BE & 0.040$^{+0.021}_{-0.010}$ & $0.08\pm0.03$ & &  & 
\enddata
\tablecomments{The uncertainties on all dust masses in the Table were determined from bootstrap Monte Carlo simulations of the data (see Section \ref{dustmass}). The dust masses listed for the nucleosynthesis models are the maximum allowed values assuming a 100\% condensation efficiency in the ejecta, and the yields for a 11 $\rm M_{\odot}$ progenitor from \citet{wos95} (WW95), and a 13 $\rm M_{\odot}$ progenitor from \citet{nom06} (N06) and \citet{wos07} (WH07).}
\end{deluxetable*}

\subsection{Dust Composition}

The mid-IR spectrum of dust in the Crab Nebula is fairly featureless \citep{tem12b}, consistent with the generally featureless spectra generated by silicate and carbon grains. Theoretical dust condensation models do indeed show that a large fraction of dust formed in Type IIp SNe is in the form of silicate and carbon \citep[e.g.][]{koz09}. 
In our models, we use three different sets of optical constants to calculate the mass absorption coefficients: those from \citet{wei01} to characterize the silicate grains; those from \citet{rou91} to characterize the amorphous carbon grains (labeled AC throughout the paper); and those from \citet{zub04} for amorphous carbon grains of type BE. The values of $\kappa$ as a function of wavelength for each of the three grain compositions is shown in Figure~\ref{qabs}.

The real and imaginary parts of the refractory index of AC carbon, $n$ and $k$, extend only to 300~\micron. They smoothly increase in the 10-300 \micron\ wavelength range, following the functional form that can be approximated by a power law $\lambda^{-0.06}$ and $\lambda^{-0.15}$ for n and k, respectively. These power law indices were used to extrapolate the optical constants out to $10^4$  \micron, leading to a smooth $\kappa$ which maintains a constant emissivity index at long wavelengths (see Figure \ref{qabs}).

While the optical constants for carbon-BE area measured beyond 500 \micron, there are no data below $\sim0.1$ \micron, where a significant fraction of the energy absorption takes place. Therefore, dust temperatures computed for carbon-BE grains of all sizes will be lower than the physical temperatures expected from PWN heating. We included them in this work only for comparison with previous work by \citet{tem12b} and \citet{gom12}. Because of the lack of short wavelength coverage for the optical constants of carbon-BE, we chose carbon-AC to represent the carbon dust in the ejecta. 

\begin{figure}
\epsscale{1.2} \plotone{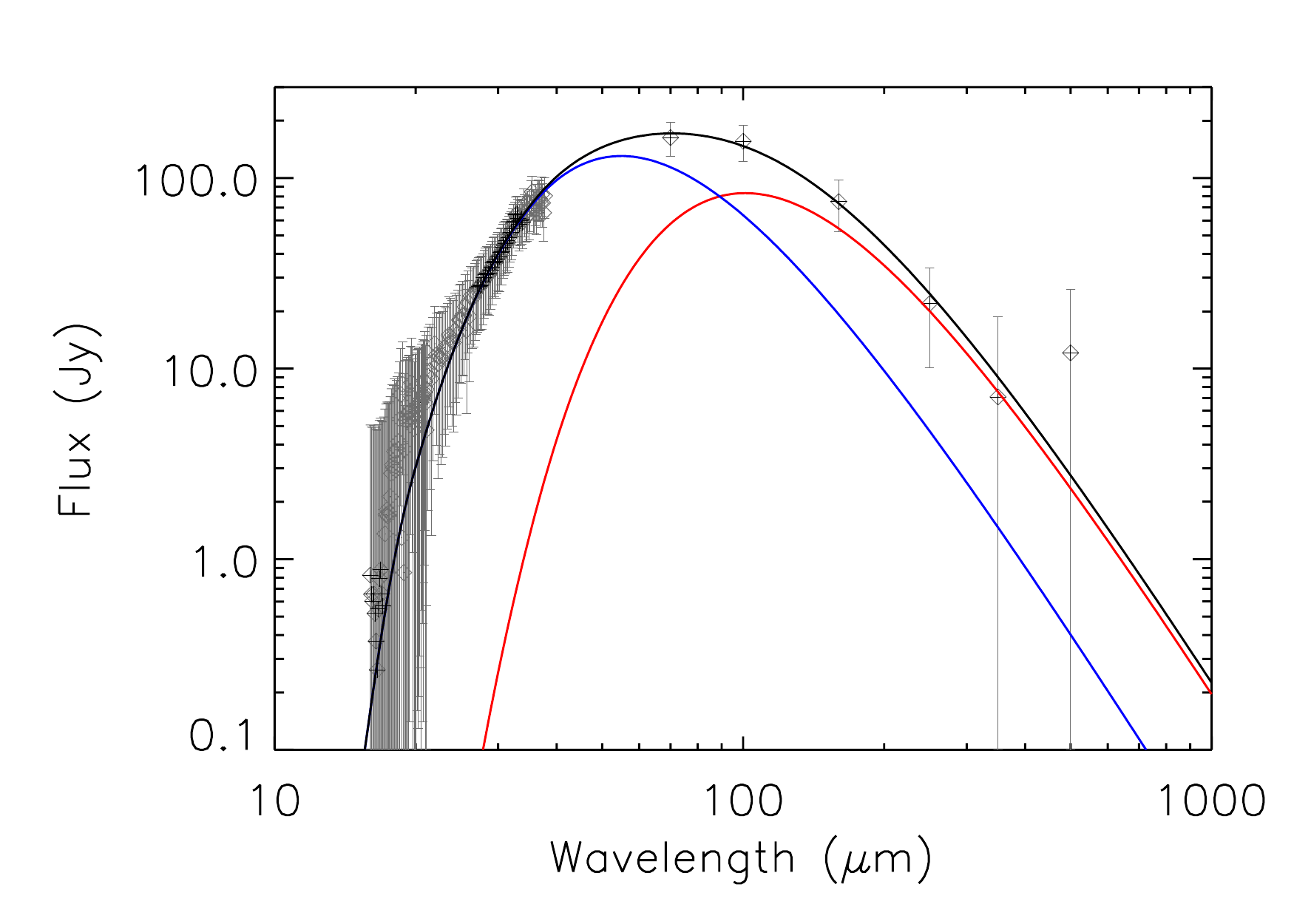}\\
\epsscale{1.2} \plotone{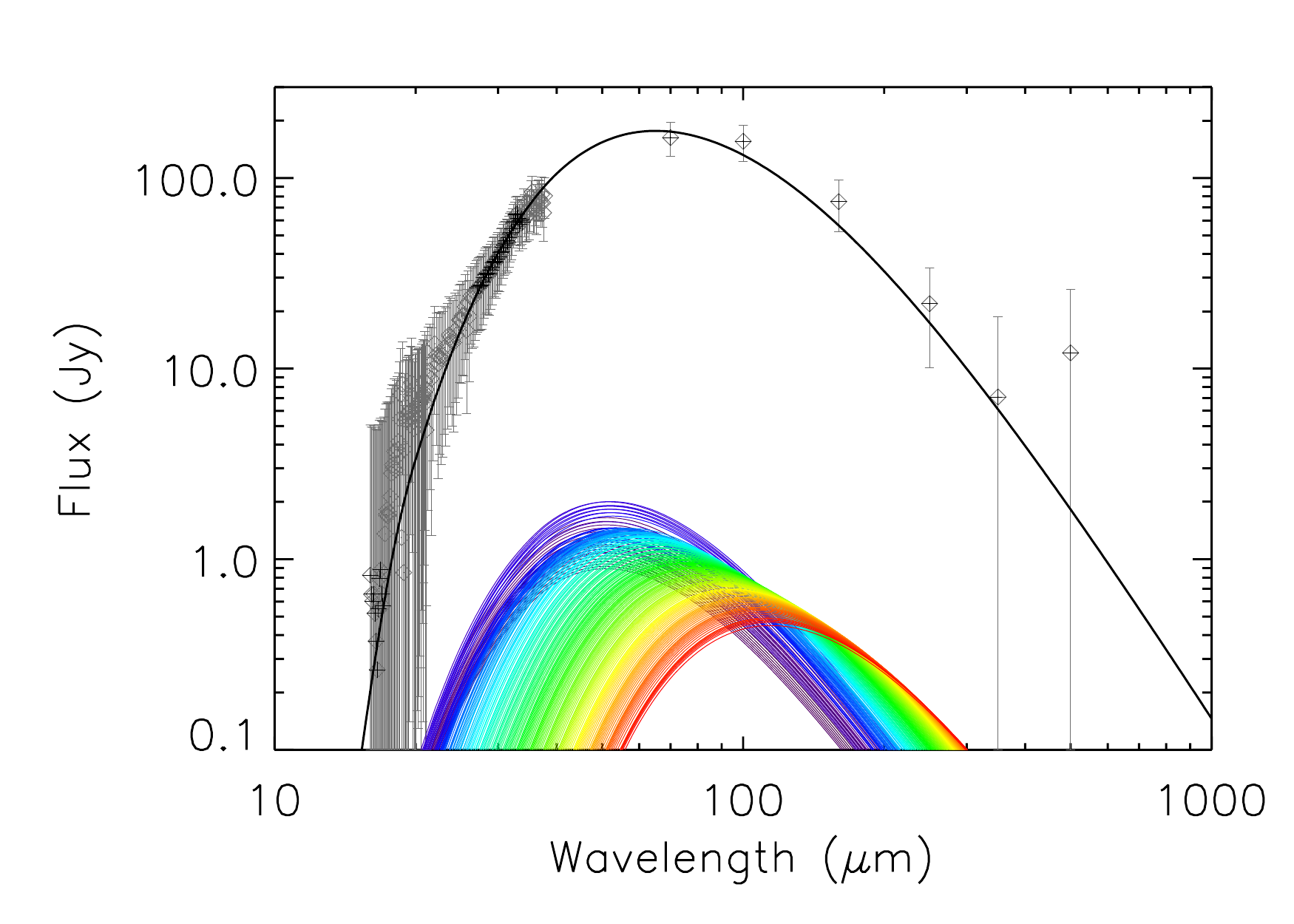}
\epsscale{1.2} \plotone{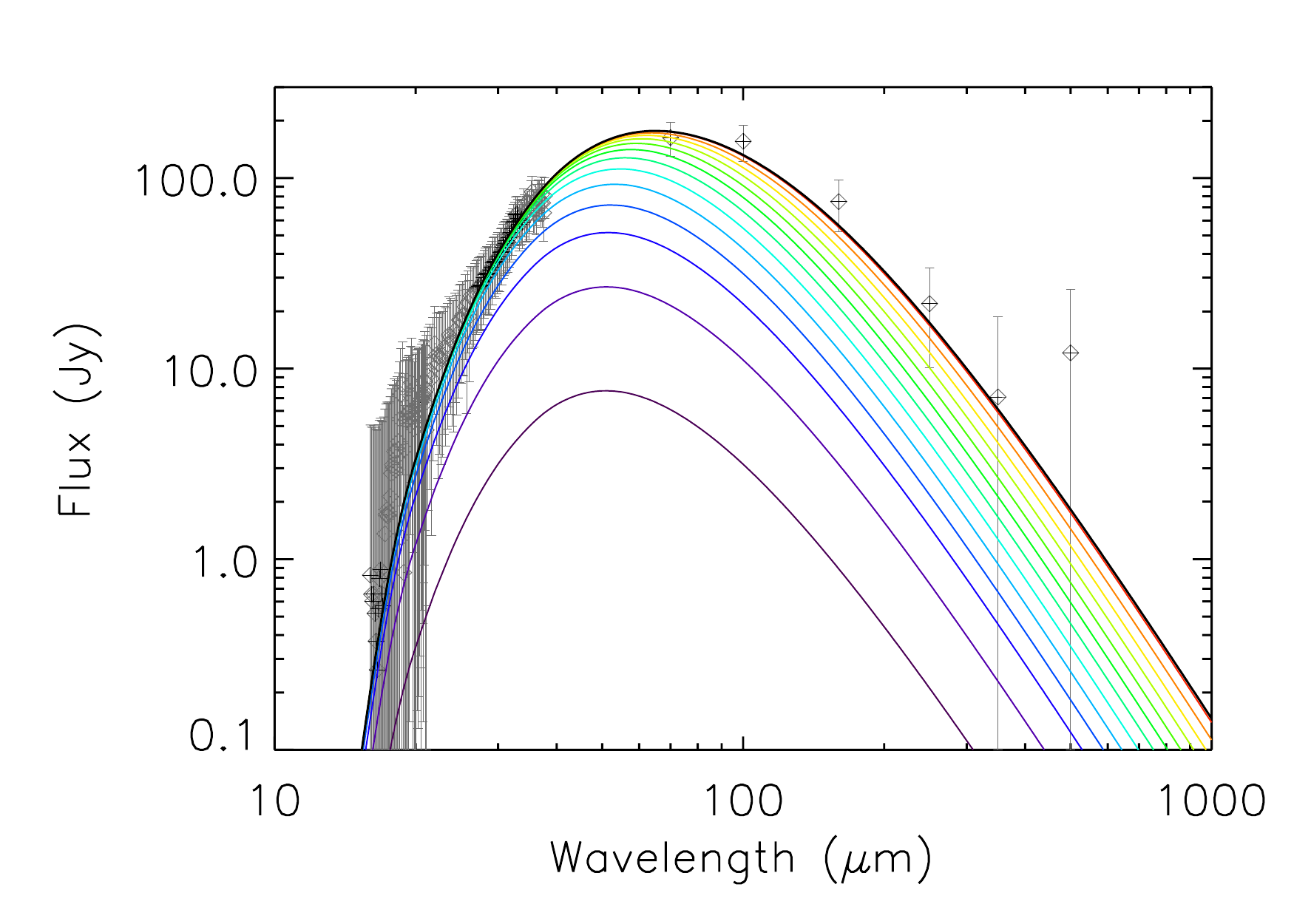}\caption{\label{silfits} \textit{Top}: Best-fit two-temperature fit for the silicate grain composition. The blue curve represents the warm grains with a temperature of 55$\pm$11 K, while the red curve represents the cold 30$\pm$5 K grains. The black curve is the sum of the individual components. \textit{Middle}: Our best-fit dust heating model for silicate grains. The rainbow colored curves represent the spectra of individual dust grains of different sizes that are heated to different temperatures. The black curve represents the sum of the individual rainbow-colored spectra. \textit{Bottom}: The same model as in the middle plot, but with the individual spectra added cumulatively starting from the hottest grains.}
\end{figure}

\begin{figure}
\epsscale{1.2} \plotone{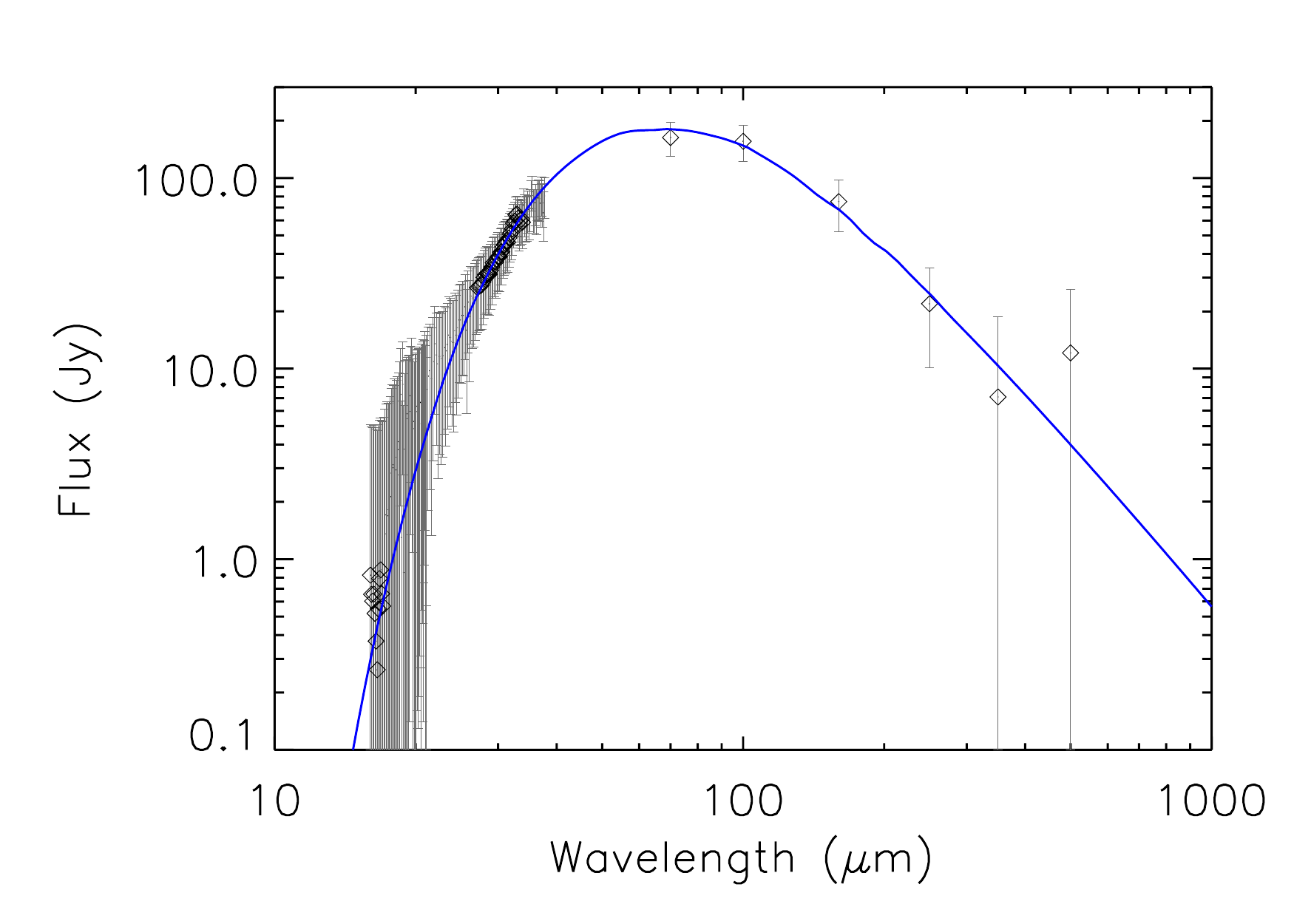}
\epsscale{1.2} \plotone{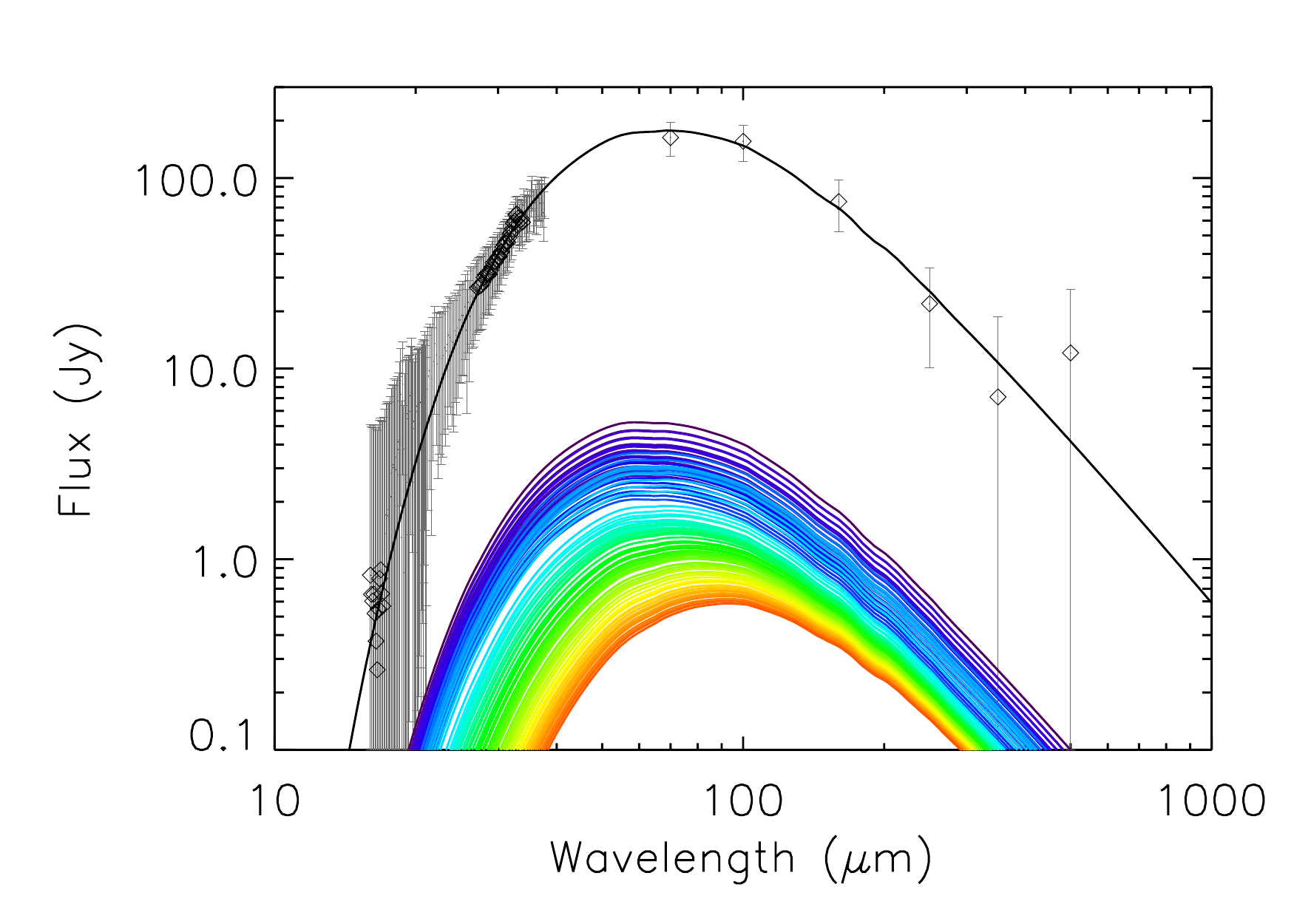}
\epsscale{1.2} \plotone{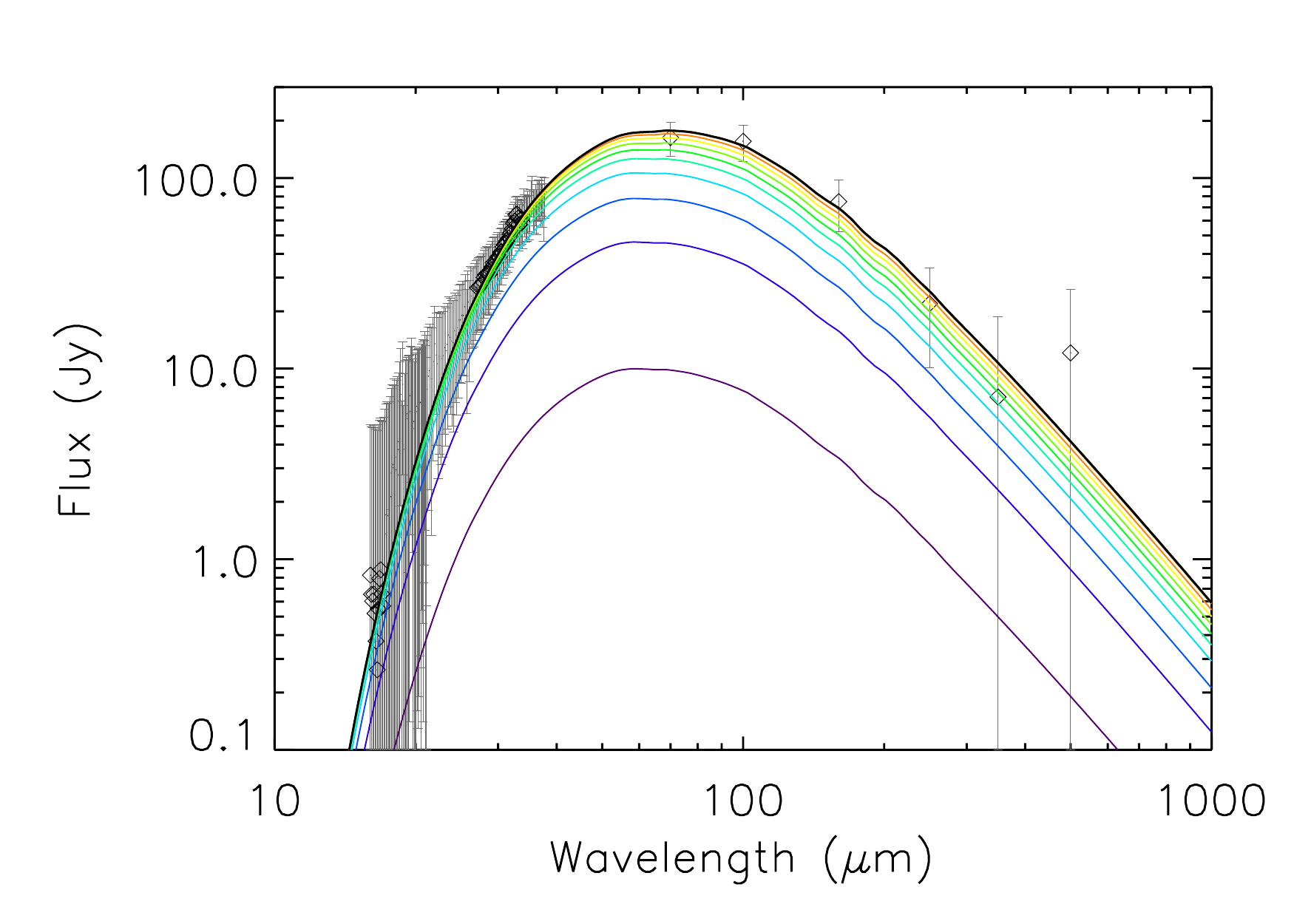}\caption{\label{crbfits} \textit{Top}: Best-fit one-temperature fit for the carbon-AC grain composition. The blue curve represents a grain temperature of 56$\pm$2 K. \textit{Middle}: Our best-fit dust heating model for carbon-AC grains. The rainbow colored curves represent the spectra of individual dust grains of different sizes that are heated to different temperatures. The black curve represents the sum of the individual rainbow-colored spectra. \textit{Bottom}: The same model as in the middle plot, but with the individual spectra added cumulatively starting from the hottest grains.}
\end{figure}

\subsection{Grain Size Distributions}

The determination of the grain size distribution in SNRs requires detailed knowledge of the physical condition in the SN ejecta, and the dust heating mechanism. The size distribution of SN condensed dust is therefore a priori unknown.
Models for the formation of dust in SNe \citep{tod01, noz10} have derived the size distribution of the various dust species that form in SN ejecta. However, these calculations assumed a uniform ejecta, and that the growth of the dust grains proceeded only through the accretion of single monomers, leaving out possible growth through coagulation in the ejecta.

Considering the above mentioned uncertainties, we adopt a general parametrization for the grain size distribution, described by a power law in grain radii, $a^{-\alpha}$, and a lower and upper grain radius cutoff on the grain radii, $a_{min}$ and $a_{max}$ respectively. Historically, such characterization was used by \citet{mrn} (MRN) to derive the size distribution of interstellar dust from the observed UV-optical extinction \citep[for a review see][]{cla03}. The incorporation of additional observational constraints: the diffuse IR emission, interstellar abundances, and diffuse interstellar scattering, have yielded a more complex interstellar grain size distribution \citep{wei01,li01,zub04}. This distribution is determined by the  size distribution of the grains that are injected from all sources into the ISM, and by the various interstellar processes that alter their sizes, including thermo-kinetic sputtering, evaporative and shattering grain-grain collisions, accretion and coagulation. Attempts to characterize the net size distribution resulting from all these processes were made by \citet{lif89, odo97, hir12}.

In order to fit the mid and far-IR dust emission in the Crab Nebula, we constructed a grid of grain size distributions with the power-law index $\alpha$ ranging from 0.0-6.0, and $a_{max}$ ranging from 0.03 to 5.0~\micron. The maximum limit of 5.0~\micron\ is already larger than the maximum grain size obtained from dust condensation models \citep[e.g.][]{tod01}, and grain sizes larger than this are not expected to form in SN ejecta.
By allowing $a_{max}$ to vary, we hope to determine to what radius SN grains can grow, and to compare the best-fit parameters to theoretical models for grain growth in Type IIP SNe. 

Fitting the minimum grain size cut-off $a_{min}$ allows us to estimate the amount of future grain processes in SNRs. However, in the Crab Nebula the reverse shock has not yet reached the PWN, so we do not expect that any grain destruction has yet occurred. We tested the effect of varying the minimum grain size on the best-fit parameters in our model by allowing $a_{min}$ to be as high as 0.03 \micron\, the lowest allowed value of $a_{max}$. Due to the lack of short-IR dust spectra, the $a_{min}$ parameter is not well constrained by our model. However, we did confirm that varying this parameter does not affect the best fit dust mass and shape of the grain size distribution, since most of the dust mass is contained in the larger grains. We therefore fixed the value of $a_{min}$ to be 0.001~\micron.

\subsection{Fitting of the IR Spectrum}\label{fitting}

We fitted our entire grid of models to the observed dust emission from the Crab Nebula that include the average \textit{Spitzer} IRS spectrum of the dust emission scaled to the total synchrotron and line-subtracted MIPS 24 \micron\ flux \citep{tem12b}, and the synchrotron and line-subtracted \textit{Herschel} PACS 70, 100, and 160 \micron\ flux measurements, SPIRE 250 and 350 \micron\ measurements, and revised \textit{Spitzer} fluxes from \citet{gom12}, where cold dust emission still contributes. Even though the line emission was subtracted from the entire IRS spectrum, our fits only included the wavelength regions of the IRS spectrum where the line emission did not contribute significantly. The data and the best-fit model for each grain composition are shown in Figure~\ref{bestfit}. The portions of the IRS spectrum data were used in the fit are overplotted in black. 

The best-fit values for the distance from the heating source $r$, power-law size distribution index $\alpha$, and the maximum grain size cut-off $a_{max}$ are summarized in Table~\ref{dustfitstab}. The dust model for the amorphous carbon-AC grain composition provides the best fit to the data, since it simultaneously provides a good fit to the mid-IR spectrum and the \textit{Herschel} data points at far-IR wavelengths. The best-fit silicate grain composition falls somewhat short at far-IR wavelengths, indicating that the presence of colder grains would be required to fit the far-IR data points. This would either require a grain radius larger than our limit of 5 \micron, or it may mean that some of the dust may be located farther away from the heating source than the best-fit average distance.

\subsection{Monte Carlo Simulations}\label{montecarlo}

We note that the large uncertainties in the data points in the IRS spectrum are dominated by systematic uncertainties introduced by the subtraction of the underlying synchrotron spectrum \citep[see][]{tem12b}. For this reason, the reduced $\chi^2$ for our best fits is too small, on the order of $\sim0.06$. Since our uncertainties are not random statistical uncertainties, we cannot use $\chi^2$ statistics in determining confidence levels for our fitted parameters. Instead, we used the bootstrap method to estimate the spread in the fitted parameters. The absolute $\chi^2$ values from our fits are used only to demonstrate the relative goodness of fit for the various models.  

In order to estimate the uncertainties in the fitted parameters while correctly accounting for the statistical uncertainties in the dust spectrum introduced by the subtraction of the synchrotron spectrum, we generated 5,000 simulated spectra using the bootstrap method. The spectra were simulated by adding an average global synchrotron spectrum to select IRS data (black data points in Figure~\ref{bestfit}), with a synchrotron power-law index of 0.42, as found by \citet{gom12}. We then produced 5,000 sample dust spectra by re-subtracting a global synchrotron spectrum while randomly selecting spectral indices within 3$\sigma$ ($\pm$0.02) of the best fit value of 0.42. We also randomly selected values for the 3.6 \micron\ flux from which the synchrotron spectrum is extrapolated in the range of 12.65$\pm$0.63 Jy \citep{tem12b}, and we accounted for the $\sim$5\% uncertainty that results from normalizing the IRS dust spectrum to the total flux from dust in the MIPS 24 \micron\ band. The \textit{Spitzer} MIPS and \textit{Herschel} data were similarly varied within the uncertainties shown in Figure~4 of \citet{gom12}. The resulting range of simulated spectra are shown as the gray band in Figure~\ref{montecarlospec}, with the actual data overlaid in black.
We then fitted our grid of dust models to each of the simulated spectra to obtain a distribution in the best-fit values for the  power-law index $\alpha$ of the grain size distribution, the value of $\rm a_{max}$, and the total dust mass for each composition. The results for each parameter will be discussed in the following sections.

\section{RESULTS AND DISCUSSION}\label{results}

\subsection{Distribution of Grain Temperatures}\label{temps}

The main difference between using a physical dust heating model and a simple two-temperature fit to the data is that in the former, the range of dust temperatures and their distribution is uniquely determined by the grain size distribution and the heating mechanism. Figure~\ref{temp} shows the dust temperature ($T$) as a function of grain size, assuming that the heating source (i.e. the Crab's PWN) is located at the best-fit distance of 0.5 pc. The colored curves show the equilibrium grain temperature as a function of grain size for each of the three grain compositions that were used in the fit. The shape of the temperature profile was obtained by equating the dust heating and cooling rates, as described in Section~\ref{heat}. An important thing to notice is the significant difference in the temperature of carbon-AC and carbon-BE grains. The additional coverage between 0.002 and 0.05 \micron\ (0.02-0.6 keV) for the carbon-AC grains raises the temperature for all grain sizes, but has a most apparent affect on the smaller grains (see Figure~\ref{temp}). While the carbon-AC grains with sizes below $\sim$0.01 \micron\ are heated to a temperature of $\sim$68 K, the same-sized carbon-BE grains only attain a temperature of $\sim$ 55 K. The large difference in dust temperatures between these two carbon compositions is not due to the differences in grain properties or the shape of the absorption coefficients $Q_{abs}$, but is instead primarily due to the limited wavelength coverage of $Q_{abs}$. A significant fraction of the radiative energy of the PWN is emitted at energies above 0.02~keV (0.05~\micron), where there are no data for the absorption efficiencies of the carbon-BE dust. Accounting for energies above 0.02 keV is especially important for heating by PWNe, since their non-thermal spectra peak at these energies. So even if we cannot determine the nature of the carbon dust in the Crab, whether it is AC or BE, our choice of carbon-AC dust is primarily motivated by the fact that it gives a more physical representation of the temperature distribution in the nebula. 

Figure~\ref{silfits} shows the comparison between a two-temperature fit to the dust emission in the Crab Nebula using a silicate grain composition (top panel), and our best-fit physical dust heating model with parameters listed in Table~\ref{dustfitstab} (middle and bottom panels). The middle panel of Figure~\ref{silfits} shows spectra of grains with different sizes as individual rainbow colored curves that all add up to the best-fit black curve. The bottom panel shows the same model, but with every tenth cumulative sum of individual spectra shown in rainbow colors. While both models provide equally good fits to the IR data, the physical dust heating model allows us to constrain physical properties of the dust, and reduces the total dust mass responsible for the IR emission (see Section~\ref{dustmass}). Figure~\ref{crbfits} shows the same three plots, but for the carbon-AC grain composition. We note that the IR spectrum can be well fitted with only a single temperature component of carbon-AC. The physical model still produces a range of temperatures, but from the comparison of Figures~\ref{silfits} and \ref{crbfits}, it is evidence that the best fit physical model produces a narrower temperature range for carbon-AC than for silicates.

\begin{figure*}
\epsscale{0.55} \plotone{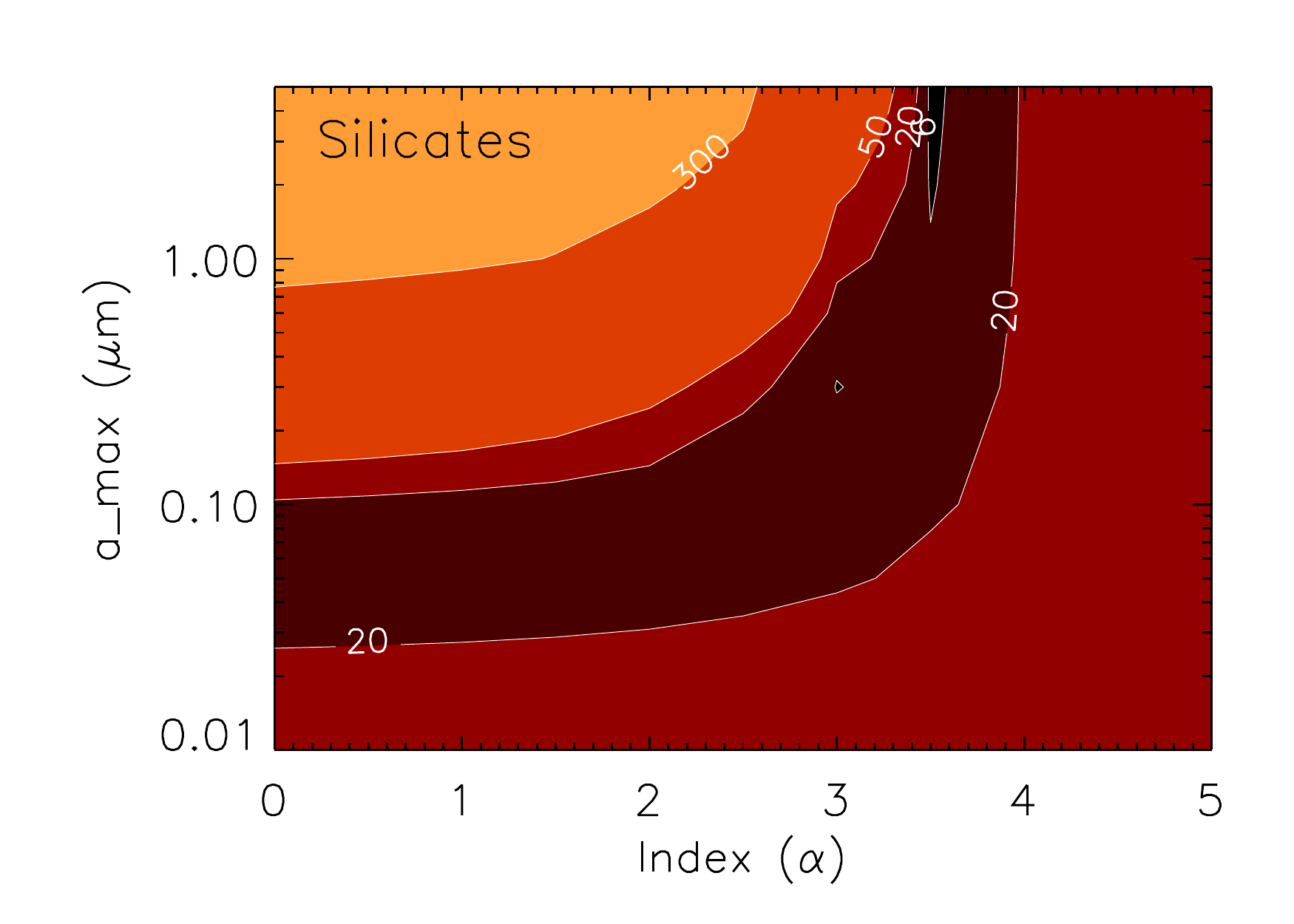}
\epsscale{0.55} \plotone{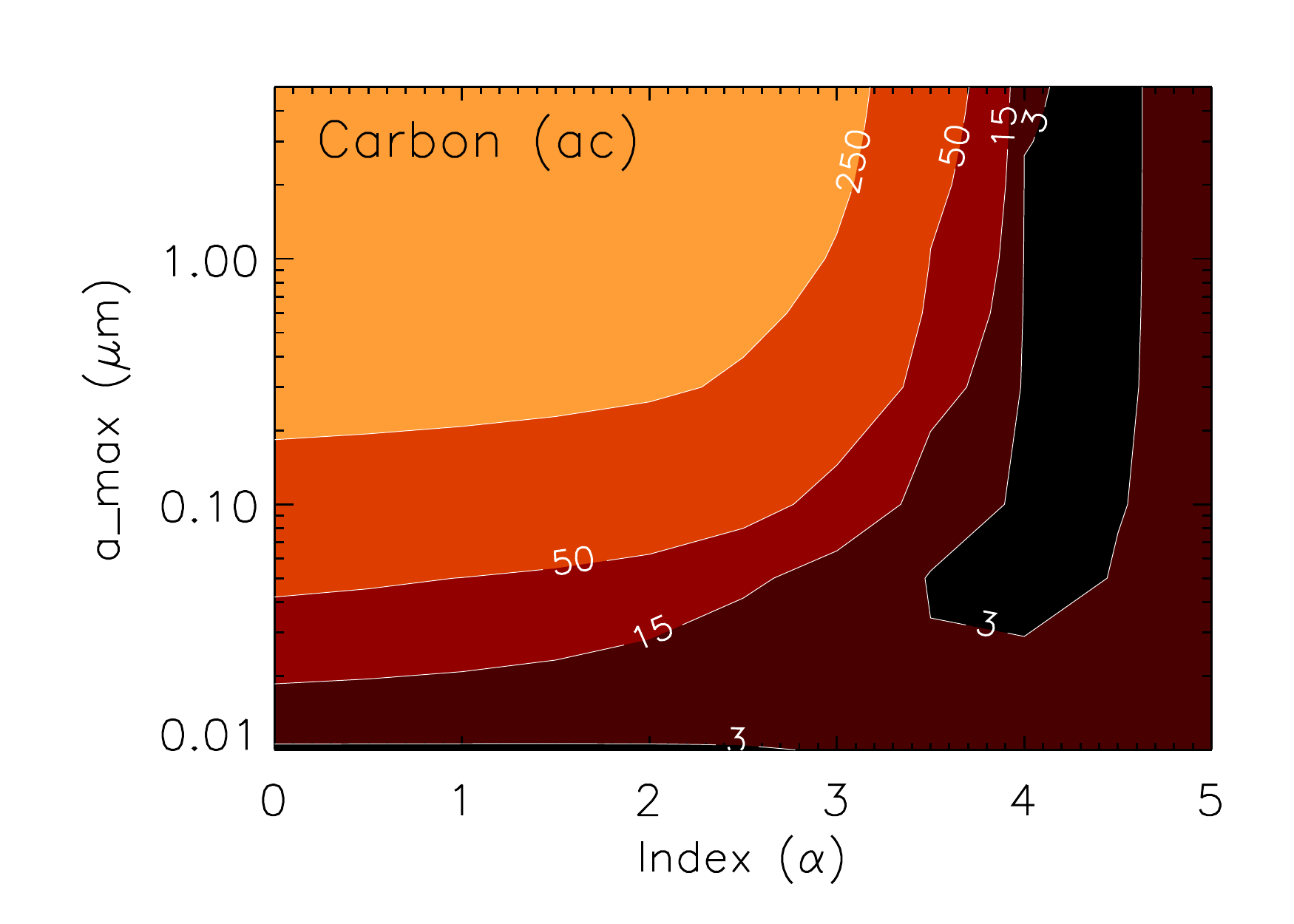}\caption{\label{chi2}$\chi^2$ contour plots for the maximum grain size cutoff, $\rm a_{max}$, and the index on the power-law grain size distribution $\alpha$. The contours indicate that an $\alpha$ value of 3.5-4.0 and a larger grain size cutoff tend to produce better fits for both silicate and carbon-AC grain compositions.}
\end{figure*}

\subsection{Grain Size Distribution}

The dust heating model also allows us to place some constrains on the shape of the grain size distribution and the maximum size of grains that formed in the Crab Nebula. Figure~\ref{chi2} shows the $\chi^2$ contour plots for $\alpha$, the index of the power-law size distribution, versus the maximum grain size cut-off $a_{max}$. The parameter values for models with minimum $\chi^2$ values are listed in Table~\ref{dustfitstab}. The contour plots show the relative goodness of fit for the range of parameter values used in our grid of dust distribution models. For all compositions, the best-fit value for the power-law index is $\alpha$=3.5-4.0. The fits to the 5,000 Monte Carlo simulations show that the power-law index is very well constrained with a standard deviation of only 0.1. This size distribution is similar to what is found for the ISM \citep[e.g.][]{mrn}. Physical modeling of dust emission for a larger sample of SNRs is needed to determine if similar size distributions are found in other core-collapse SNe.

In \citet{tem12b}, we found that the dust emission in the Crab observed by \textit{Spitzer} implies a small grain size of $<$ 0.05 \micron. However, since we only had coverage up to 70 \micron, we were sampling only the smaller grains that were heated to temperatures of around 50 K (see Figure~\ref{temp}). The addition of the \textit{Herschel} long wavelength data, and a better constrains on the synchrotron index derived from the \textit{Planck} data \citep{gom12}, has now allowed us to place some constraints on the maximum size of dust grains in the Crab's filaments. Our best-fit model appears to favor a fairly large maximum grain size cut-off (see Figure~\ref{chi2}). The value for $a_{max}$ for the best-fit model is 0.1 \micron\ or larger, depending on the grain composition (Table~\ref{dustfitstab}). This is consistent with models of dust formation in Type IIP SNe \citep{koz09,noz10}. These models suggest that the dust mass of grains formed in Type IIP SNe with massive H-envelopes is dominated by grains larger than 0.03 \micron, while the mass of dust for Type IIb SNe is dominated by very small grain $<$ 0.006 \micron. Probing the grain size in SNRs through physical heating models can therefore be a useful tool for comparing and constraining dust condensation models. 
The Crab Nebula is thought to be a result of a Type IIP explosion, primarily based on the low expansion velocities of the ejecta filaments and the low progenitor mass \citep[e.g.][]{che05,mac08}. The large dust grain radii inferred from our models are consistent with a Type IIP origin. We note that the larger grain size also implies that a large fraction of the dust mass will survive the eventual reverse shock interaction \citep{dwe05,koz09,bia09,noz10}.

\begin{figure*}
\epsscale{0.55} \plotone{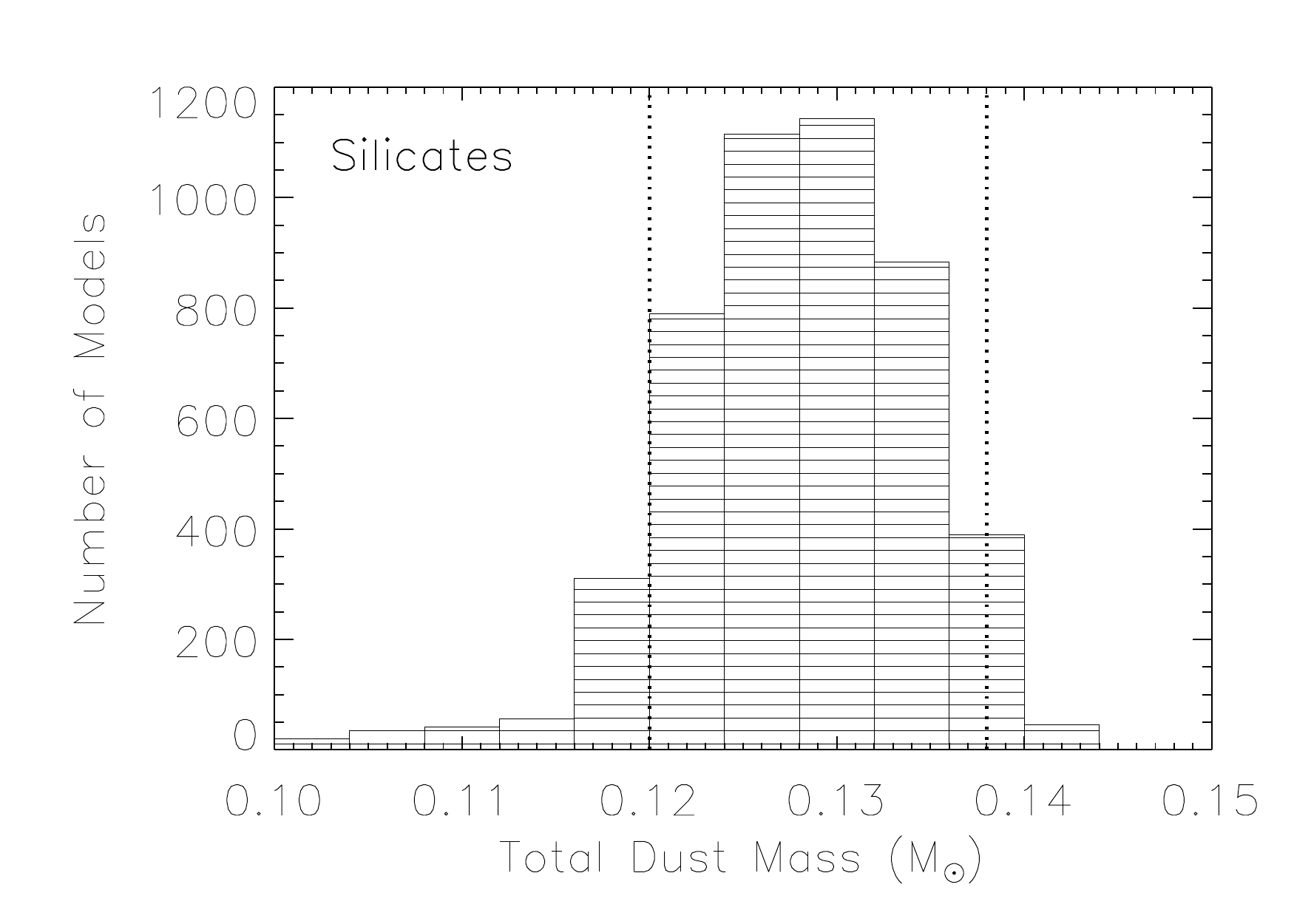}
\epsscale{0.55} \plotone{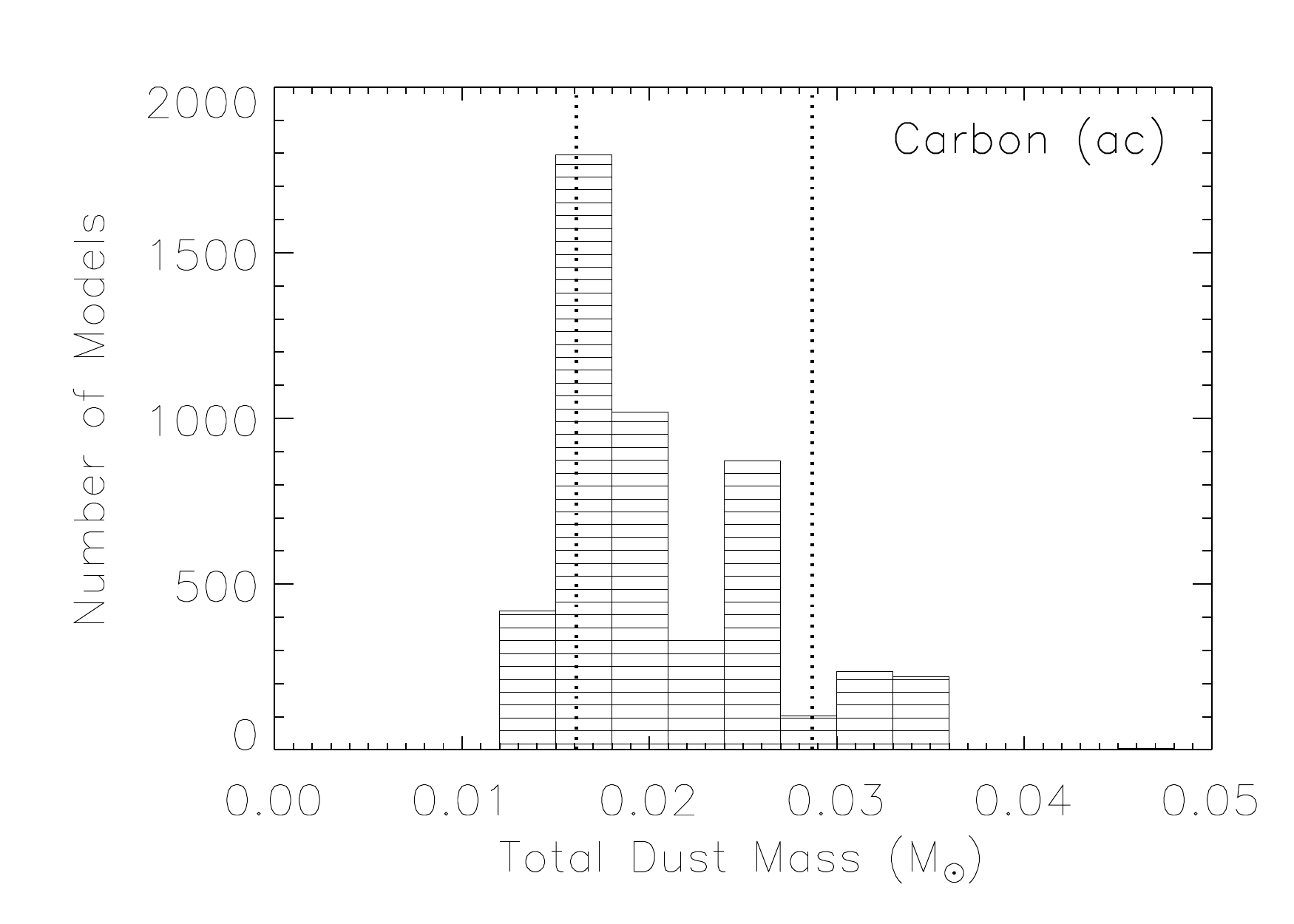}
\caption{\label{mass}Distribution of total dust masses obtained from the bootstrap fits to the 5000 simulated spectra using the silicate (top) and carbon-AC grains (bottom). The dotted vertical lines represent limits for which 90\% of masses lie either above or below the corresponding mass. These mass limits are taken as the uncertainty range in our best-fit total masses (see Table~\ref{dustmasstab}).}
\end{figure*}

\subsection{Revised Dust Mass for the Crab Nebula}\label{dustmass}

In order to compare our derived dust masses to previous studies, we first reproduced the two-temperature fits from \citet{gom12} with the same dust grain compositions that they used in their analysis. Our best-fit temperatures are consistent with the values of \citet{gom12}. The total masses that we derive are $0.2\pm0.1\rm \: M_{\odot}$ and $0.08\pm0.03\rm \: M_{\odot}$ for silicate and carbon-BE grains, respectively. They are listed in Table~\ref{dustmasstab}, and also found to be consistent with the values from \citet{gom12} of $0.25^{+0.32}_{-0.08}\rm \: M_{\odot}$ for silicates and $0.12\pm0.01\rm \: M_{\odot}$ for carbon-BE grains.

We then calculated the total dust mass corresponding to our best-fit dust heating models characterized by a grain size and temperature distribution, the latter obtained by calculating the temperature of each dust grain as it is radiatively heated by the PWN.  We find that the total mass of the silicate and carbon-BE dust is reduced by a factor of $\sim$2. We derive a dust mass of  $0.13\pm0.01\rm \: M_{\odot}$ for silicates and $0.04^{+0.02}_{-0.01}\rm \: M_{\odot}$ for carbon-BE grains. The total mass for the carbon-BE model is presented only to demonstrate the effect of using a continuous temperature distribution, as opposed to a bimodal temperature distribution, on the calculated dust mass. As discussed in previous sections, the lack of short-wavelength data for optical constants for carbon-BE grains prevents us from estimating the actual grain temperature for this composition. Consequently, the resulting dust properties are not a physical representation of their actual values in the Crab Nebula. For this reason, we chose to model the carbon emission using carbon-AC dust, which produces a more realistic distribution of dust temperatures for the PWN-heated dust. 

The total dust mass derived from our model for carbon-AC is $0.019^{+0.010}_{-0.003}\rm \: M_{\odot}$. The difference in the mass between the two different carbon compositions is in part due to the difference in their optical constants at long wavelengths. However, the comparison between the masses derived for these two different compositions is unphysical, since the lack of optical data for the carbon-BE dust at X-ray and UV wavelengths significantly affects the derived temperature distribution, and hence the mass.
Surprisingly, we find that the IR data are well fitted even with a single temperature component of carbon-AC dust with the same mass as derived from our physical model and a temperature of $T = 56\pm2 \:\rm K$ (see Table~\ref{dustfitstab} and Figure~\ref{crbfits}). This shows that our parametrization of the grain size distribution is flexible enough to produce a temperature distribution leading to the same dust mass as that derived from a simple single-temperature fit. It also suggests that a distribution of grain temperatures does not automatically lead to a significant change in dust mass and that, for a given IR spectrum, the magnitude of the effect depends on the dust composition.

Table~\ref{dustmasstab} also lists the uncertainties in the total masses that were derived using the bootstrap method described in Section~\ref{montecarlo}. 
Figure~\ref{mass} shows the histograms of the total dust mass corresponding to the best-fit model for each of the 5,000 simulated spectra. The dotted vertical lines represent limits for which 90\% of masses lie either above or below the corresponding mass. We take the uncertainty range in our derived masses to be the range between these two limits, and we list them in Table~\ref{dustmasstab}. 
The mass uncertainties in our two-component fits were also determined by the bootstrap method. The comparison between the mass uncertainties shows that the total mass determined by the physical model is better constrained than the the simplified one and two temperature models, for both silicate and carbon grain compositions.
Overall, the strict limits placed on the temperature distribution by our PWN heating model allow for a more physical and accurate determination of the total mass. The absolute uncertainties on the total mass are likely larger due to the uncertainty on the distance to the Crab Nebula, however, this is a systematic uncertainty that will equally affect both the physical and two-temperature mass estimates.

Carbon-AC dust provides a somewhat better fit to the IR emission in the Crab Nebula than silicate dust, suggesting that the dust mass may either be dominated by carbon dust, or may possibly be a mixture of the two dust compositions. The best-fit silicate model falls short at longer wavelengths and would require an unrealistically large grain size (larger than our limit of 5 \micron) to reach cold enough temperatures to better fit the long wavelength data. A second dust component is therefore required to provide a better fit to the data, suggesting that the composition is more likely to be a mixture of carbon and silicate dust.
This  is also supported by the fact that the derived mass of silicate dust is in excess of the total mass of available refractory elements needed to condense these grains. Table~\ref{dustmasstab} lists the the nucleosynthetic yields from core-collapse SNe for progenitors in the 11-13 $\rm \: M_{\odot}$ range \citep{wos95,wos07,nom06}. We assumed a 100\% grain condensation efficiency in the ejecta, and calculated the upper limit on the carbon and silicate dust yield from the total yield of carbon, oxygen, magnesium, silicon, and iron. To obtain the upper limit on the silicate grain mass, we summed the maximum possible masses for $\rm SiO_2$, $\rm MgO$, and $\rm Fe_{3}O_{4}$ grains. The results show that the dust masses derived from the two-temperature component models are clearly too large, especially considering that the progenitor of the Crab Nebula is estimated to be around $\sim9.5\rm \: M_{\odot}$ \citep{mac08}. The total mass of silicate dust derived from the physical models is also too large compared to most nucleosynthesis yields, suggesting that the dust in the Crab Nebula is predominantly in the form of carbon dust.

\section{CONCLUSIONS}

We modeled the mid and far-IR dust emission from the Crab Nebula in order to determine the grain properties and total dust mass in the ejecta. We use a physical dust heating model in which a continuous power-law size distribution of dust grains is radiatively heated by the PWN, giving rise to the thermal IR continuum. The best-fit models to the dust emission favor a grain size distribution with a power-law index of 3.5-4.0, and a relatively large maximum grain size cut-off of $>$ 0.1 \micron, consistent with theoretical predictions of dust formation in Type IIP SNe \citep{koz09,noz10}. 

The IR emission is best described by an amorphous carbon grain composition with a total mass of $0.019^{+0.010}_{-0.003}\rm \: M_{\odot}$, or a silicate grain composition with a mass of $0.13\pm0.01\rm \: M_{\odot}$. The dust is likely to be a mixture of carbon and silicate grains, with a total mass between these two values. Both values are lower than previous estimates from more simplified one or two-temperature models. The difference in the dust mass is due to the use of a continuous grain size and temperature distribution derived from a physical model, and a different choice of dust compositions with a more complete coverage of their optical constants over the range of energies at which the heating by the PWN is important.  

The limits placed on the temperature distribution by our physical heating model allow for a more accurate determination of the total mass. Even though a distribution of grain temperatures may not necessarily lead to a significant change in dust mass, as is the case for carbon-AC, modeling the data with a physical distribution of temperatures is required to determine the magnitude of the effect.

In recent years, most large dust mass estimates for SNRs assumed the simplified one or two-temperature models, including the mass estimates from \textit{Herschel} data for Cas A, SN 1987A, and the Crab Nebula \citep[e.g.][]{bar10,sib10,mat11,gom12}. Our modeling results imply that a physical dust emission model is important for deriving a the actual mass of dust in SN ejecta.

Such models will also provide information on the size distribution of the grains, which is important for estimating the ultimate survival of the dust following the encounter with the reverse shock. 
 The amount of SN-condensed dust in the Crab Nebula is significantly lower than that required to produce the massive amount of dust observed in high-redshift galaxies even if the dust is not destroyed in the ISM.

The importance of using physical models to determine dust properties in SN ejecta is not limited to cases of radiatively heated dust. Modeling of collisionally-heated dust in, for example, SN1987A \citep{dwe10}, Puppis~A \citep{are10}, Kes~75 \citep{tem12}, or LMC remnants \citep{wil11}, provide, in addition to dust masses and grain size distributions, also information on grain sputtering efficiencies in the X-ray emitting plasmas, as well as a diagnostics of the plasma conditions in the shocked gas. 
Physical models are thus crucial for deriving dust masses and dust properties for other PWNe and SNRs. Such studies are important for determining the role of SNe as dust sources in the local and high-redshift universe.

\acknowledgements{We thank Rick Arendt and George Sonneborn for useful discussions and comments on the paper, and an anonymous referee for his/her comments which led to valuable improvements in the paper.}

\bibliographystyle{apj}


\end{document}